\documentclass{svmult}
\usepackage{amssymb,amsmath,amsfonts}
\usepackage[utf8]{inputenc}
\usepackage{float}
\usepackage{bm}
\usepackage{amsmath}
\usepackage{amssymb}
\usepackage{float}
\usepackage{graphicx}
\usepackage{soul}
\usepackage{color}
\usepackage{lineno}
\newcommand{\new}[1]{\textcolor{black}{#1}}
\newcommand{\old}[1]{\textcolor{black}{#1}}

\begin{document}

\title{Navigation and control of outcomes in a generalized Lotka-Volterra model of the microbiome}

\titlerunning{Navigation and control of outcomes in a gLV model of the microbiome}

\author{Eric W. Jones \and Parker Shankin-Clarke \and Jean M. Carlson}

\institute{Department of Physics, University of California at Santa Barbara, Santa Barbara CA 93106, USA, \texttt{ewj@physics.ucsb.edu}}

\maketitle

\abstract{
The generalized Lotka-Volterra (gLV) equations model the microbiome as a
collection of interacting ecological species.  Here we use a particular
experimentally-derived gLV model of {\em C. difficile} infection (CDI) as a
case study to generate methods that are applicable to generic gLV models.  We
examine how to transition gLV systems between multiple steady states through
the application of direct control protocols, which alter the state of the
system via the instantaneous addition or subtraction of microbial species.  
%In the CDI model, the state of the system is more easily altered by direct control protocols that introduce a foreign population composed of many species (e.g. a steady state composition of the system) rather than protocols that are made up of a single microbial species.
Then, the geometry of the basins of attraction of point attractors is
compressed into an {\em attractor network}, which decomposes a multistable
high-dimensional landscape into web of bistable subsystems.  This attractor
network is used to identify efficient (total intervention volume
minimizing) protocols that drive the system from one basin to another.  
In some cases, 
the most efficient
control protocol is circuitous and will take the system through intermediate
steady states with sequential interventions.  Clinically, the efficient control of the microbiome has pertinent
applications for bacteriotherapies, which seek to remedy microbiome-affiliated
diseases by directly altering the composition of the gut microbiome. 
%We use the dimensionality-reduction technique steady-state reduction to analytically (and therefore efficiently) generate approximate attractor networks.
%These attractor networks demonstrate how the state of a system can be driven between basins of attraction via direct interventions.
%including the treatment of  conditions as diverse as \textit{Clostridioides difficile (C. difficile)} infection and autism spectrum disorder via bacteriotherapies that 
%Taking gLV models as a proxy for the true underlying microbial dynamics, these results indicate the potential efficiacy of sequentially-administered bacteriotherapies.
%We draw an analogy between these driving an ecological model with instantaneous addition of microbes and driving the underlying physical system with bacteriotherapies.
%Thus, by using gLV models as a proxy for true microbial dynamics, our work might inform how to improve the administration of bacteriotherapies.
%One method of modeling the microbiome is with the generalized Lotka–Volterra (gLV) equations, in which microbial dynamics are generated by a set of coupled second-order ordinary differential equations.
%For non-periodic dynamics landscape (or Lyapunov function) with geometry dictated by the parameters of the model.
%The human microbiome consists of the bacteria that reside in the human gastrointerstinal tract.
}

\setlength{\parindent}{5ex}

\section{Introduction}\label{sec:intro}
In this chapter we characterize the dynamics of the generalized Lotka-Volterra
(gLV) equations, a set of nonlinear coupled differential equations that are
traditionally used in theoretical ecology to study interacting populations.  In
particular, gLV models are examined in the context of the gut microbiome, which
consists of an ensemble of microorganisms that inhabit the gastrointestinal
tract.  Our scope is restricted to ecological dynamics that relax towards
attractors; in this case microbial dynamics follow a pseudo-energy landscape
similar to a Lyapunov function, in which minima of the landscape correspond to
steady states of the system.

Canonically, landscape-based descriptions of
biological processes have been used to describe how cell fates are determined
in Waddington's epigenetic landscape, and more recently they have been employed
to study the genetic landscape of gene regulatory networks
\cite{huang2009cancer}.  In this gene regulatory network study, high-dimensional dynamics 
are characterized using an {\em attractor network},
which compresses the geometric structure of the basins of attraction of fixed
points into a
graph. In this chapter we apply attractor networks to describe
a multi-stable gLV system by a web of interconnected bistable landscapes.  By
mapping the structure of dynamical landscapes, attractor networks inform the
control of steady-state outcomes in gLV systems.

The study of the microbiome is motivated by a desire to better understand the ecological dynamics that underlie microbiome behavior, which might advance the ability of clinicians to respond to microbiome-associated disorders and to improve microbiome health.
Experimental evidence linking microbiome to host health is reviewed in Section \ref{subsec:microbiome}, and alternative mathematical models of the microbiome are summarized in Section \ref{subsec:modeling}.

As a case study, we consider an experimentally-derived gLV model of {\em Clostridiodes difficile} infection (CDI), and use attractor networks to inform how to navigate between stable microbial communities of the system.
The system is manipulated with {\em direct interventions} that modify an existing microbial state by either introducing a foreign microbial population (referred to as a {\em transplant}) or by removing existing microbial species.
These direct interventions are interpreted as numerical implementations of bacteriotherapies,  and when they are successful these interventions drive a microbial state into the basin of attraction of a target state. 
%construction to inform how to optimally navigate between stable microbial communities.
%that exhibit experimentally meaningful steady states in the context of CDI.
%In the CDI model, these bacteriotherapies are most effective when the transplant population consists of many microbial species.
%Additionally, we find that sequentially-administered bacteriotherapies can take advantage of inherent ecological dynamics to be more efficient, in the sense that a smaller transplant volume is required to transition between basins of attraction.
%administration of FMT using requires a series of sequential transplantations, which theoretically corresponds to optimally navigating high dimensional ecological state space through a series of intermediary steady states.
%Attractor networks provide a framework for the direct control of gLV systems.
Broadly, these results examine the ecological mechanisms that underlie the successful administration of bacteriotherapies, and inform how the intrinsic ecological dynamics of the microbiome might be harnessed to improve the efficacy of bacteriotherapies.

%%%%%%%%%%%%%%%%%%%%%%%%%%%%%%%%%%%%%%%%%%%%%%%%%%%%%%%%
\section{Background}
%%%%%%%%%%%%%%%%%%%%%%%%%%%%%%%%%%%%%%%%%%%%%%%%%%%%%%%%

%%%%%%%%%%%%%%%%%%%%%%%%%%%%%%%%%%%%%%%%%%%%%%%%%%%%%%%%
\subsection{Generalized Lotka-Volterra (gLV) models}
%%%%%%%%%%%%%%%%%%%%%%%%%%%%%%%%%%%%%%%%%%%%%%%%%%%%%%%%

%%%%%%%%%%%%%%%%%%%%%%%%%%%%%%%%%%%%%%%%%%%%%%%%%%%%%%%%
\begin{figure}[t]
\begin{center}
\includegraphics[width=.65\textwidth]{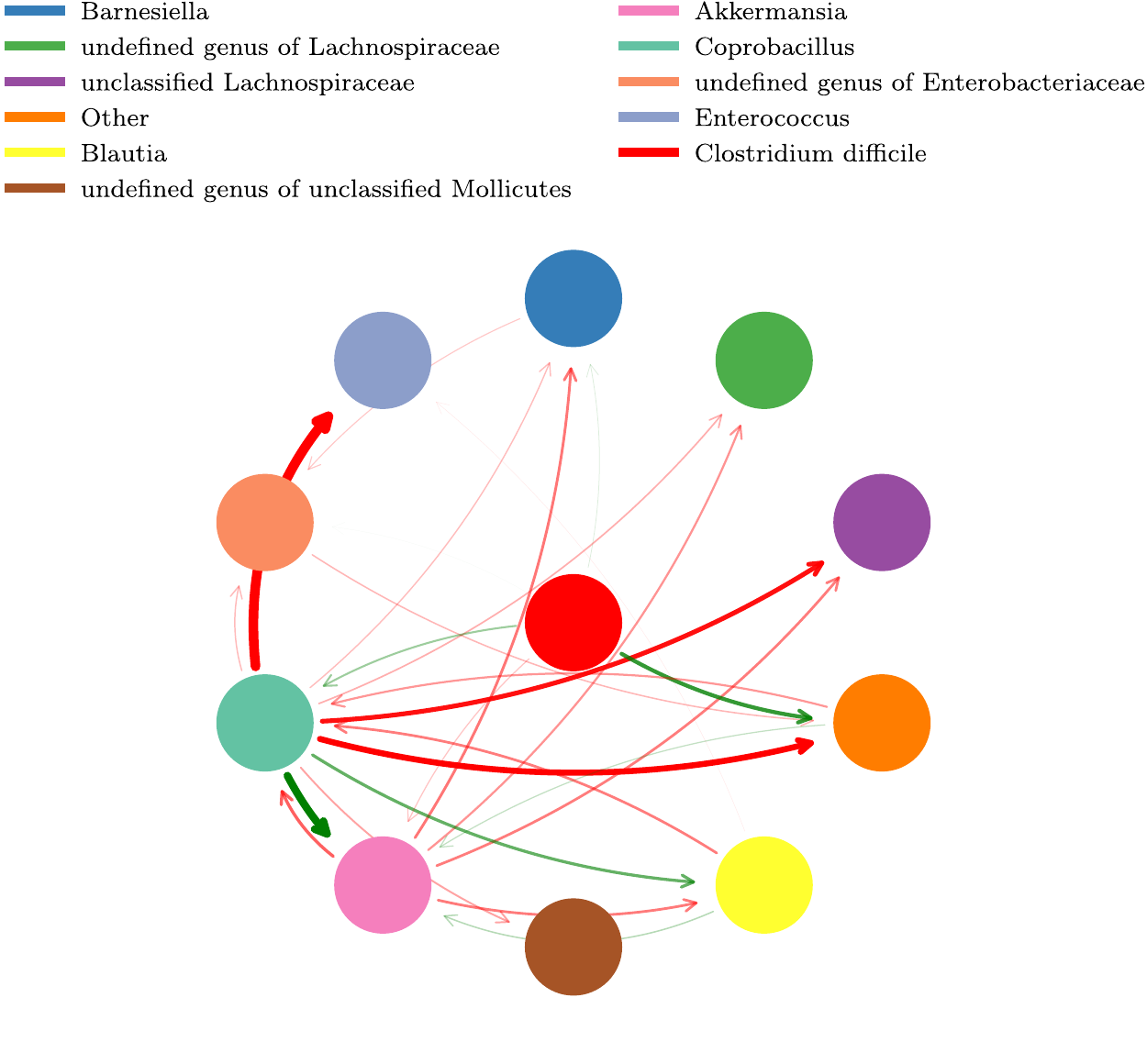}
\caption{
\textbf{Pairwise interactions between bacterial populations interpreted as a microbial food web}.
An arrow from population $j$ to population $i$ represents the effect of $j$ on the growth of $i$, as described by the interaction term $K_{ij}$ in a generalized Lotka-Volterra model Eq~(\ref{eq:gLV}).
The width and opacity of an arrow are proportional to $|K_{ij}|$, and positive interactions ($K_{ij} > 0$) are green while inhibitory interactions ($K_{ij} < 0$) are red.
The pairwise interactions $K_{ij}$ here were fit from an experimental mouse model of {\em C. difficile} infection (CDI) \cite{stein2013ecological,BuffiePamer2012}. 
%To reduce dimensionality, bacterial species of the same genus are consolidated into one population; the exception is \textit{C. difficile} (CD), which is a single bacterial species.
{\em C. difficile}, the culprit behind CDI, is colored red and located in the center of the food web.
This figure and caption are adapted from \cite{jones2018silico}. 
\label{fig:food_web}}
\end{center} 
\end{figure}
%%%%%%%%%%%%%%%%%%%%%%%%%%%%%%%%%%%%%%%%%%%%%%%%%%%%%%%%

In this chapter the generalized Lotka-Volterra (gLV) equations are treated as a mathematical proxy for the microbial dynamics of the gut microbiome.
The gLV equations describe the dynamics of a microbial population $y_i$ in a system of $N$ total interacting populations as 
%a set of autonomous, nonlinear, coupled, first-order ordinary differential equations 
\begin{equation} 
\frac{dy_{i}}{dt} = y_{i} \left( \rho_{i} + \sum_{j=1}^{N} K_{ij}y_{j} \right), \quad i \in 1,\ \ldots,\ N,
\label{eq:gLV} \end{equation}
where the growth rate of species $i$ is given by $\rho_i$, and the pairwise interaction $K_{ij}$ encodes the ecological effects of species $j$ on species $i$.
 
%The gLV equations generate species-level dynamics $x_{i}(t)$ based on growth rates $\mu_i$ of each species $i$, and interaction parameters $M_{ij}$ between species $i$ and $j$.
%The interaction parameters $M_{ij}$ encode prototypical ecological behaviors including competition, amensalism, and commensalism. 

A gLV system with $N$ populations can exhibit up to $2^N$ steady states, where each steady state is specific to a distinct presence/absence combination of the $N$ species.
%Steady states of a gLV system are interpreted as stable microbiome compositions observed in individuals.
Motivated by clinical bacteriotherapies like fecal microbiota transplantation (FMT) that seek to drive a diseased microbiome towards a healthy composition, we consider how to control the steady-state outcomes of gLV systems with direct interventions that instantaneously supplement or deplete the microbial state of a system.
%The steady states generated by the gLV model are the temporally fixed bacterial populations of a given microbiome.
%Every N bacterial species produces 2N steady states, where each state is an overlapping subset of a single steady state.
%The biological relevance of the gLV system is reflected through the inclusion of both diseased C. difficile infected states corresponding to dysbiosis and healthy states  comprised of beneficial flora corresponding to a normal microbiome. 
Towards this end, as in previous work \cite{stein2013ecological, jones2018silico} the gLV model is extended to include the instantaneous addition of a foreign microbial transplant \textbf{v} at time $t^*$, as well as the administration of an antibiotic treatment $u(t)$, so that 
\begin{equation} \frac{dy_{i}}{dt} = y_{i} \left( \rho_{i} + \sum_{j=1}^{N} K_{ij}y_{j} \right) + v_i\delta(t - t^*) + u(t) \varepsilon_i y_i, \label{eq:transplant} \end{equation}
where $v_i$ is the $i$th component of the foreign transplant \textbf{v}, $\delta(\cdot)$ is the Dirac delta function, and $\varepsilon_i$ is the antibiotic susceptibility of population $i$.
%As in previous work, we construe this instantaneous addition as a medical intervention.

Three types of bacteriotherapy-inspired direct interventions are examined, corresponding to three types of transplant $\textbf{v}$: the addition of a single species (interpreted as a probiotic), the addition of a stable external microbial community (interpreted as fecal microbial transplant), and the removal of a particular species (interpreted as a phage therapy).
More explicitly, we consider the direct control problem in which an initial condition is in the basin of attraction of some attractor 
$\mathbf{y_a}$, and the goal is to identify the smallest intervention $\mathbf{v}$ that is able to drive the system into the basin of attraction of some other target state $\mathbf{y_b}$.

\new{
We choose the cost function $\varphi$ to be the size of the transplant, 
\begin{equation}
    \varphi = \lVert \textbf{v} \rVert_1,
\end{equation}
so that the optimal control protocol minimizes the required
transplant size. 
Since the transplant inputs are instantaneous (i.e. Dirac-delta type), this is
a sampled-data control problem \cite{BourdinTrelat2016}.
}
%%%%%%%%%%%%%%%%%%%%%%%%%%%%%%%%%%%%%%%%%%%%%%%%%%%%%%%%
\subsection{Experimentally-derived gLV model of {\em C. difficile} infection}
%%%%%%%%%%%%%%%%%%%%%%%%%%%%%%%%%%%%%%%%%%%%%%%%%%%%%%%%

We use a gLV model of \textit{C. difficile} infection (CDI) as a case study for how to transition between basins of attraction.
In this model, the growth rates $\bm{\rho}$ and interaction parameters $\mathbf{K}$ were fit by Stein {\em et al.} to microbial abundance time-series data from a mouse experiment performed by Buffie {\em et al.} \cite{stein2013ecological, BuffiePamer2012}.
To reduce the number of parameters of the model, Stein \textit{et al.} assumed that bacteria within a given genus behave similarly, and consolidated the species-level data into genus-level data.
In Fig.~\ref{fig:food_web} the interaction parameters are displayed as a food web where the circles are microbial populations, and the edges describe the effect of one population on another, where positive interactions ($K_{ij} > 0$) are black and inhibitory interactions ($K_{ji} < 0$) are red.
\old{The interactions between populations have no clear hierarchy, indicating that the fitted model describes microbes on the same trophic level competing for shared resources.}
The fitted antibiotic susceptibilities $\varepsilon_i$ are mostly negative, indicating that antibiotics tend to deplete the growth rates microbial populations.

\old{The CDI model produces microbial trajectories that allow for the simulated application of medical interventions.}
Previous work explored the effects of antibiotics on microbial dynamics \cite{stein2013ecological,jones2018silico} in this CDI model, and found that
\old{the CDI model exhibits the clinically- and experimentally-observed outcome that antibiotic-treated microbiomes were vulnerable to CDI.}

%%%%%%%%%%%%%%%%%%%%%%%%%%%%%%%%%%%%%%%%%%%%%%%%%%%%%%%%
\begin{figure}[t]
\begin{center}
\includegraphics[width=\textwidth]{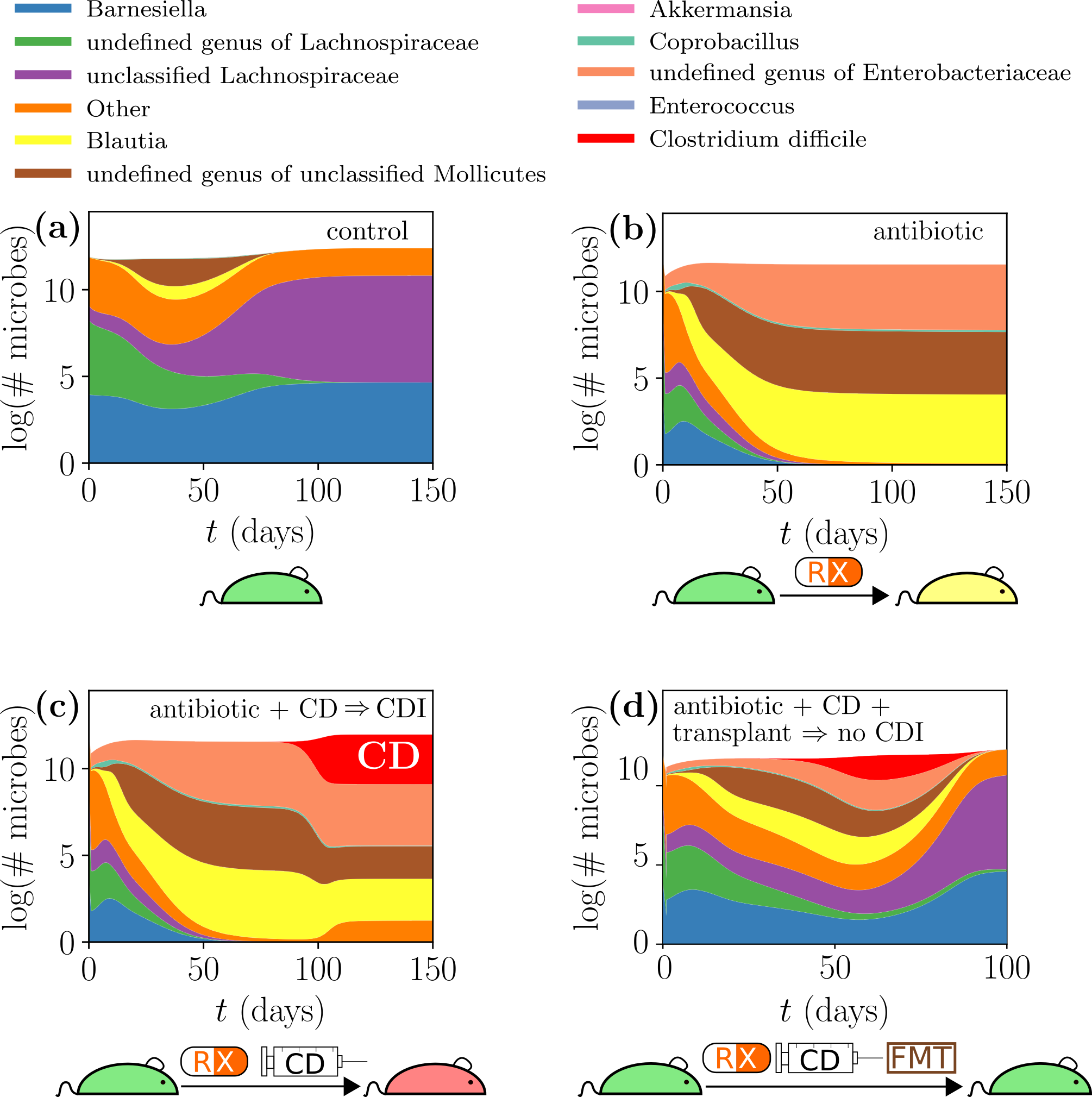}
\caption{
\textbf{External interventions can alter the steady-state outcome of an initial condition.}
All panels originate from the same experimentally-measured initial condition, but different panels correspond to different interventions: (a) no interventions occur; (b) one dose of antibiotics administered at day 0; (c) one dose of antibiotic administered at day 0, and CD inoculation on day 10; and (d) one dose of antibiotic administered at day 0, CD inoculation on day 1, followed by the immediate introduction of a foreign microbial population made up of a stable microbial community on day 1.
Together these panels demonstrate that antibiotic-induced CDI may be remedied by the administration of a direct intervention, as in fecal microbiota transplantation.
%The growth of \textit{C. difficile} (colored red) is encouraged by antibiotic treatment, since the antibiotics deplete the other microbes to a level at which \textit{C. difficile} gains a foothold.
This figure and caption are adapted from \cite{jones2018silico}. 
\label{fig:reachable_ss}}
\end{center} 
\end{figure}
%%%%%%%%%%%%%%%%%%%%%%%%%%%%%%%%%%%%%%%%%%%%%%%%%%%%%%%%

To demonstrate antibiotic-induced CDI in the gLV model, Fig.~\ref{fig:reachable_ss} shows simulated microbial trajectories that result from applying four separate intervention scenarios to an initial condition measured by Buffie {\em et al.}
\old{For clarity, these figures plot the total microbe count on a log scale (in which the total microbe count is the sum of all of the microbes in each microbial population), and at each time each microbial population is linearly colored according to its proportion at that time.} 
First, in Fig.~\ref{fig:reachable_ss}a the system is not perturbed and the system evolves to a ``healthy'' steady state, in the sense that CD is unable to invade this steady state.
In Fig.~\ref{fig:reachable_ss}b the system is exposed to a unit dose of antibiotics (i.e., $u(t) = 1$ for $0 \leq t \leq 1$, and $u(t) = 0$ otherwise) which drives the system to an ``antibiotic-depleted'' state, in the sense that CD is able to invade it.
Bearing this out, in Fig.~\ref{fig:reachable_ss}c the system is exposed to a unit dose of antibiotics and then inoculated with CD on day 10 (i.e. the transplant $\textbf{v}$ is a unit vector of CD applied at time $t^*=10$), and this system evolves towards an ``CD infected'' steady state in which CD is present.
Finally, in Fig.~\ref{fig:reachable_ss}d the 
system is exposed to a unit dose of antibiotics, inoculated with CD on day 1, and then also supplemented with an external foreign transplant (i.e. the foreign population $\textbf{v}$ is composed of an experimentally-measured initial condition introduced at a time $t^*=1$); due to the direct intervention, this system now evolves towards the healthy steady state.
%Depictions of these interventions (i.e., antibiotic application or CD exposure) are expicitly shown in Fig.~\ref{fig:reachable_ss}

\begin{figure}[t]
\begin{center}
\includegraphics[width=.65\textwidth]{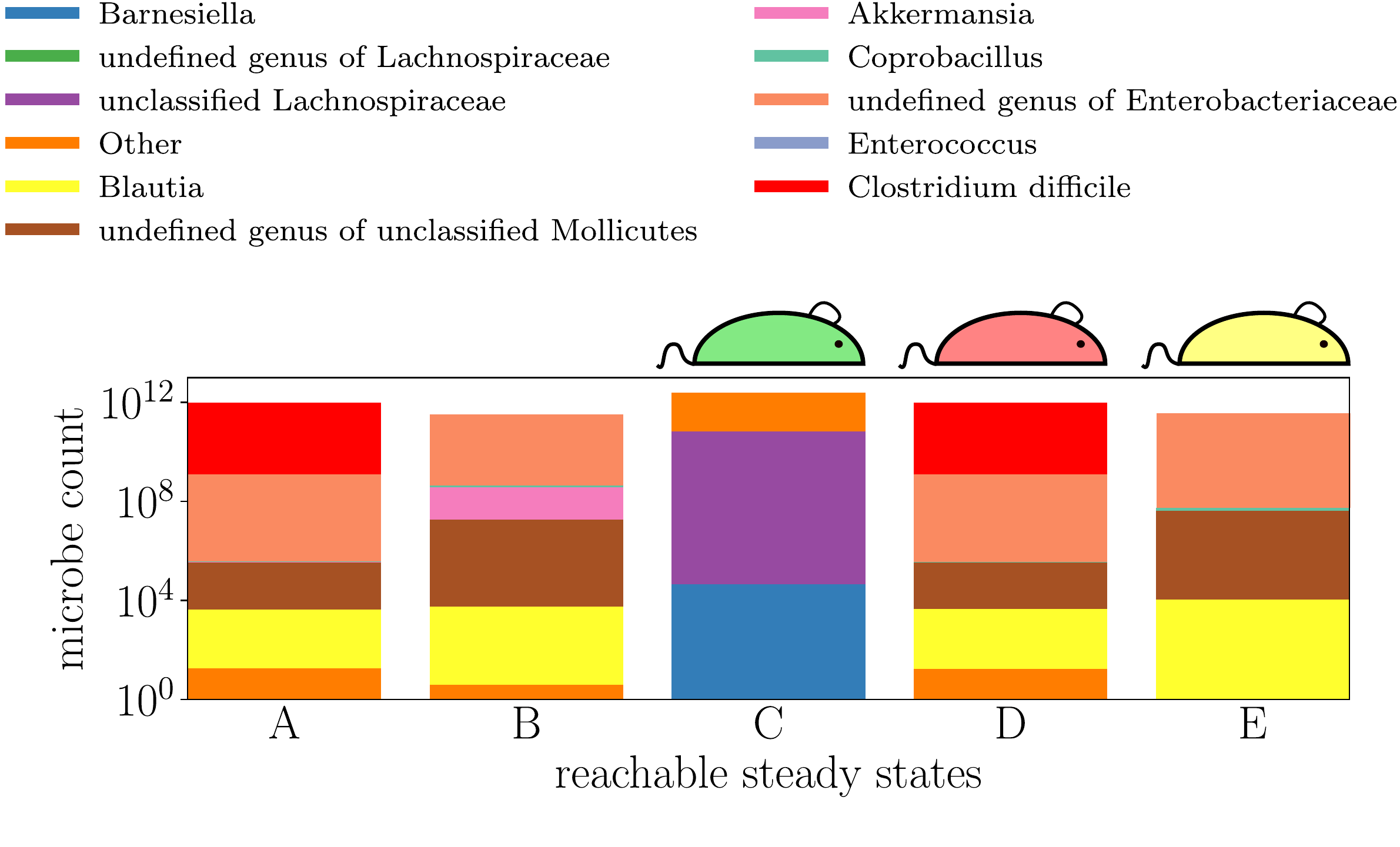}
\caption{{\bf Microbial composition of reachable steady states.}
Under the gLV model Eq.~(\ref{eq:gLV}) and for each of the nine
    experimentally-measured initial conditions, every treatment scenario tested
    in this chapter resulted in one of the steady states A-E.  Together these
    five steady states encompass a region of phase space that is relevant for
    systems originating near the nine measured initial conditions.  \new{The green,
    red, and yellow mice represent respectively the healthy, CD-infected, and
    antibiotic-depleted steady states of the CDI model.}
%To find which steady state a given treatment scenario causes, refer to Fig~\ref{fig:c_diff_phase}.
Note that while steady states A and D appear indistinguishable in this plot, their compositions vary slightly.
%The microbial compositions of each steady state are explicitly given in Table B of Supp.
This figure and caption are adapted from \cite{jones2018silico}. 
\label{fig:reachable}}
\end{center} 
\end{figure}

In each panel of Fig.~\ref{fig:reachable_ss} the colored mice represent attained steady-state microbiome compositions: green represents healthy, yellow represents antibiotic-depleted, and red represents CD-infected.
These three steady states of the CDI model resemble the compositions of the experimentally-observed mouse microbiome compositions, including their susceptibility or resilience to CD exposure \cite{BuffiePamer2012, stein2013ecological}.
In addition to these three steady states, two other steady states can be reached by initializing the system at one of the nine experimentally-measured initial conditions and then applying some type of intervention.
We call this set of five steady states the ``reachable'' steady states of the CDI model, and focus on the ecological dynamics near them.
The microbial compositions of these five steady states are displayed in Fig.~\ref{fig:reachable}. 
In this figure, the previously-mentioned healthy state is labeled steady state C, the antibiotic-depleted state is labeled steady state E, and the CD-infected state is labeled steady state D.

This experimentally-characterized CDI model provides a clinically-motivated case study in which different steady states can be associated with biologically meaningful microbiome compositions.
It is pertinent to be able to efficiently switch basins of attraction in order to attain a ``healthy'' state, and in the remainder of this chapter we investigate this goal in detail.

%%%%%%%%%%%%%%%%%%%%%%%%%%%%%%%%%%%%%%%%%%%%%%%%%%%%%%%%
%%%%%%%%%%%%%%%%%%%%%%%%%%%%%%%%%%%%%%%%%%%%%%%%%%%%%%%%

%%%%%%%%%%%%%%%%%%%%%%%%%%%%%%%%%%%%%%%%%%%%%%%%%%%%%%%%
\subsection{Approximation of bistable gLV dynamics}
%%%%%%%%%%%%%%%%%%%%%%%%%%%%%%%%%%%%%%%%%%%%%%%%%%%%%%%%

%%%%%%%%%%%%%%%%%%%%%%%%%%%%%%%%%%%%%%%%%%%%%%%%%%%%%%%%
\begin{figure}[t]
\begin{center}
\includegraphics[width=.65\textwidth]{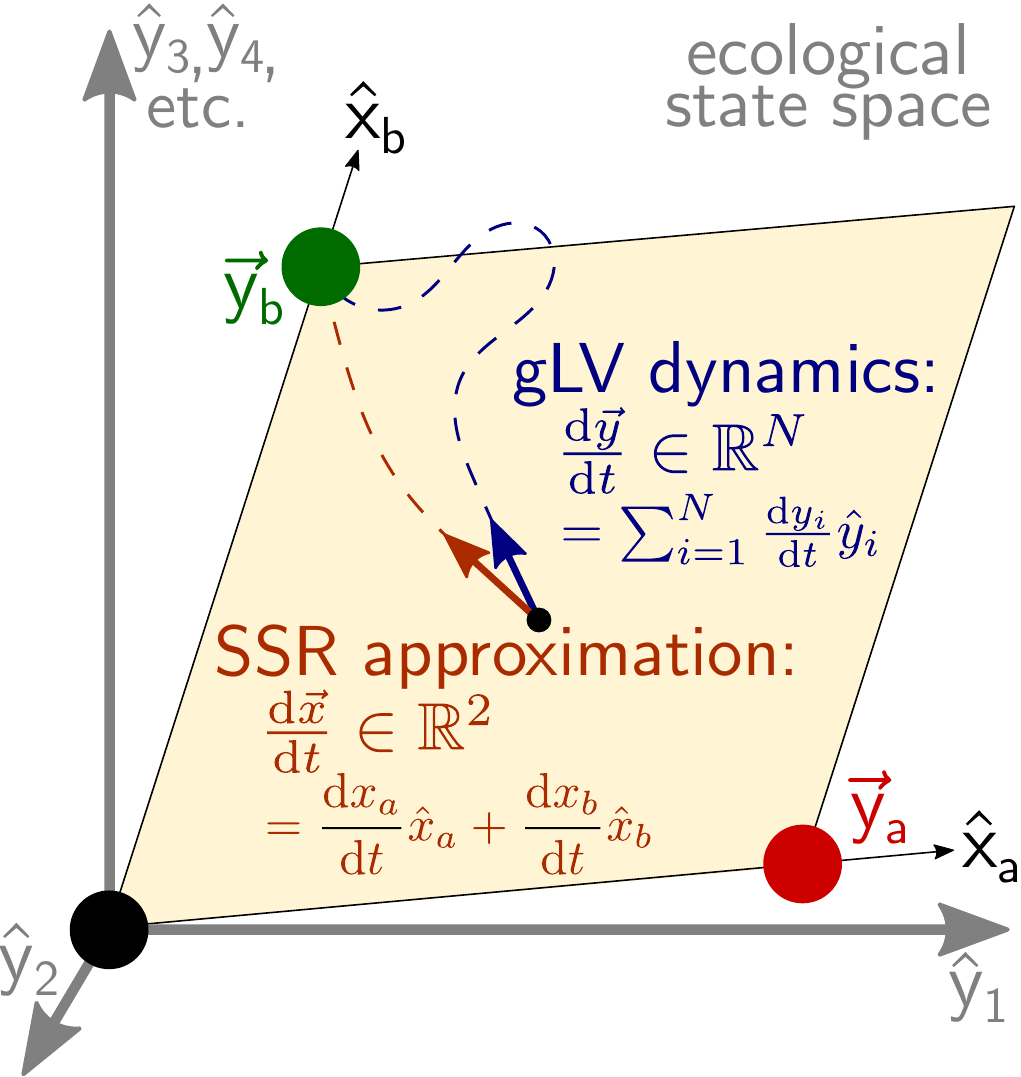}
\caption{{\bf Schematic of steady-state reduction (SSR).}
A gLV system of $N$ species (Eq.~(\ref{eq:gLV})) exhibits two steady states $\vec{y}_a$ and $\vec{y}_b$, characterized as diseased (red) and healthy (green).
SSR identifies the two-dimensional (2D) gLV system defined on the 2D subspace spanned by the two high-dimensional steady states (Eq.~(\ref{eq:gLV_2D})) that best approximates the high-dimensional system.
Specifically, SSR prescribes 2D parameters (Eqs.~(\ref{eq:ssr_params}) and (\ref{eq:ssr_params2})) that minimize the deviation between the N-dimensional gLV dynamics $\text{d}\vec{y}/\text{d}t$ and the embedded 2D SSR-reduced dynamics $\text{d}\vec{x}/\text{d}t$.
This figure and caption are adapted from \cite{JonesCarlson2019}. 
\label{fig:SSR}}
\end{center} 
\end{figure}
%%%%%%%%%%%%%%%%%%%%%%%%%%%%%%%%%%%%%%%%%%%%%%%%%%%%%%%%

The ecological dynamics of high-dimensional gLV systems are straightforward to simulate, but difficult to investigate analytically.
%Here, all simulations were produced with native ODE solvers in \texttt{python}.
Often the most fundamental features of a model are captured not by the
particular high-dimensional vector that describes the state of a system, but rather by an abstract biological outcome associated with the state of a system.
For example, in the case of the CDI model, the most important question is whether the system will tend towards a healthy (steady state C) or an antibiotic-depleted state (steady state E), and the precise composition of those steady states is not as important.
Thus, in some sense there is a low-dimensional description of the transition
between a pair of steady states, even while the actual ecological dynamics flow in a high-dimensional state space.

In earlier work we exploited this abstract outcome-oriented perspective to design steady-state reduction (SSR) \cite{JonesCarlson2019}.
\old{This method compresses a bistable region of a high-dimensional gLV model
into a reduced two-state gLV model in which the two basis populations
correspond to the bistable states of the original model.}
\old{
As depicted in Fig.~\ref{fig:SSR}, this reduced two-dimensional (2D) gLV model is defined on the 2D subspace spanned by a pair of steady states of the original model, and the subspace itself is embedded within the high-dimensional ecological phase space of the original gLV model.
The parameters of the reduced model are weighted combinations of the parameters of the original model, with weights that are related to the composition of the two high-dimensional steady states.
Within this subspace, these reduced dynamics constitute the best possible 2D gLV approximation of the high-dimensional gLV dynamics.
}
Though gLV systems in general are capable of displaying periodic or chaotic
behaviors, here our attention is restricted to trajectories that approach a
fixed point, and in particular to regions of phase space that are near the border of the basins of attraction of a pair of steady states (i.e., near the separatrix).

More explicitly, to determine the SSR-reduced 2D gLV system associated with a bistable high-dimensional gLV model, first
\old{define variables $x_a$ and $x_b$ in the direction of unit vectors $\hat{x}_a$ and $\hat{x}_b$ that parallel the two steady states according to $\hat{x}_a \equiv \vec{y}_a/\lVert \vec{y}_a \rVert_2$, and $\hat{x}_b \equiv \vec{y}_b/\lVert \vec{y}_b \rVert_2$, where $\lVert \cdot \rVert_2$ is the $2$-norm.
The 2D gLV dynamics on the subspace spanned by $\hat{x}_a$ and $\hat{x}_b$ are given by 
\begin{align} \begin{split} \frac{\text{d}x_a}{\text{d}t} &= x_a (\mu_a + M_{aa}x_a + M_{ab} x_b), \quad \text{and} \\ \frac{\text{d}x_b}{\text{d}t} &= x_b (\mu_b + M_{ba}x_a + M_{bb}x_b).
\label{eq:gLV_2D} \end{split} \end{align}
Here, the \textit{in-plane dynamics} on this subspace in vector form are written $\frac{\text{d}\vec{x}}{\text{d}t} = \frac{\text{d}x_a}{\text{d}t}\hat{x}_a + \frac{\text{d}x_b}{\text{d}t}\hat{x}_b$.
}

\old{The 2D parameters generated by SSR are chosen to minimize the deviation between the in-plane and high-dimensional gLV dynamics $\epsilon = \lVert \frac{\text{d}\vec{y}}{\text{d}t} - \frac{\text{d}\vec{x}}{\text{d}t} \rVert_2$ for any point on the subspace spanned by $\hat{x}_a$ and $\hat{x}_b$.}
The values of the SSR parameters that minimize $\epsilon$ are derived in \cite{JonesCarlson2019}.
When the two steady states $\vec y_a$ and $\vec y_b$ are orthogonal, the 2D parameters are given by
\begin{eqnarray} 
    \mu_\gamma &= \frac{\vec{\rho}\cdot \vec{y}_\gamma^{\circ 2}}{\left
    \lVert\vec{y}_\gamma\right\rVert^2_2}, \text{ and} \nonumber \\
    M_{\gamma\delta} &= \frac{(\vec{y}_\gamma^{\circ 2})^TK\vec{y}_\delta}
    {\left\lVert\vec{y}_\gamma\right\rVert^2_2\left\lVert\vec{y}_\delta
    \right\rVert_2} \label{eq:ssr_params},
\end{eqnarray}
where $\vec{y}^{\circ 2} \equiv \text{diag}(\vec{y})\vec{y}$ is the  element-wise square of $\vec{y}$.
When $\vec y_a$ and $\vec y_b$ are not orthogonal the interspecies interaction terms become slightly more complicated and are given by
\begin{eqnarray}
    M_{ab} &=
    \frac{\sum_{i,j=1}^NK_{ij}(y_{ai}y_{bj}+y_{bi}y_{aj})
    (y_{ai}-y_{bi}\sum_{k=1}^N y_{ak}y_{bk})}
    {1-(\sum_{i=1}^Ny_{ai}y_{bi})^2},
    \nonumber \text{ and} \\
    M_{ba} &=
    \frac{\sum_{i,j=1}^NK_{ij}(y_{bi}y_{aj}+y_{ai}y_{bj})
    (y_{bi}-y_{ai}\sum_{k=1}^N y_{bk}y_{ak})}
    {1-(\sum_{i=1}^Ny_{bi}y_{ai})^2}, \label{eq:ssr_params2}
\end{eqnarray}
where $\gamma,\delta \in a,b$, and $y_{ai}$ and $y_{bi}$ are the $i$th
components of the of the unit vectors $\hat{y}_a \equiv \vec{y}_a / \lVert
\vec{y}_a \rVert_2$ and $\hat{y}_b \equiv \vec{y}_b / \lVert
\vec{y}_b \rVert_2$, respectively.
\old{
Under this construction, the high-dimensional steady states $\vec{y}_a$ and $\vec{y}_b$ have in-plane steady state counterparts at $(\lVert \vec{y}_a \rVert_2,\ 0)$ and $(0,\ \lVert \vec{y}_b \rVert_2)$, respectively.}

Crucially, this reduced 2D gLV system is analytically tractable: the separatrix can be written analytically \cite{JonesCarlson2019}, and bifurcation analyses readily inform how interaction parameters alter the basins of attraction of a system \cite{WangCarlson2020}.
SSR associates complex high-dimensional dynamics with an intuitive low-dimensional system. 
In the context of the CDI model, this SSR-based understanding informs the transition between e.g. the healthy ``green'' and antibiotic-depleted ``yellow'' microbial states, and eschews the high-dimensional details.
Later, we will decompose multistable gLV systems into a web of bistable subsystems, then use this web to describe the geometry of the basins of attraction of steady states of the model as an {\em attractor network}.
Additionally, SSR will be leveraged to analytically compute the location of the separatrix in each subsystem, which is much more efficient than computing the locations of the separatrices numerically. 

%%%%%%%%%%%%%%%%%%%%%%%%%%%%%%%%%%%%%%%%%%%%%%%%%%%%%%%%
\begin{figure}[t]
\begin{center}
\includegraphics[width=.65\textwidth]{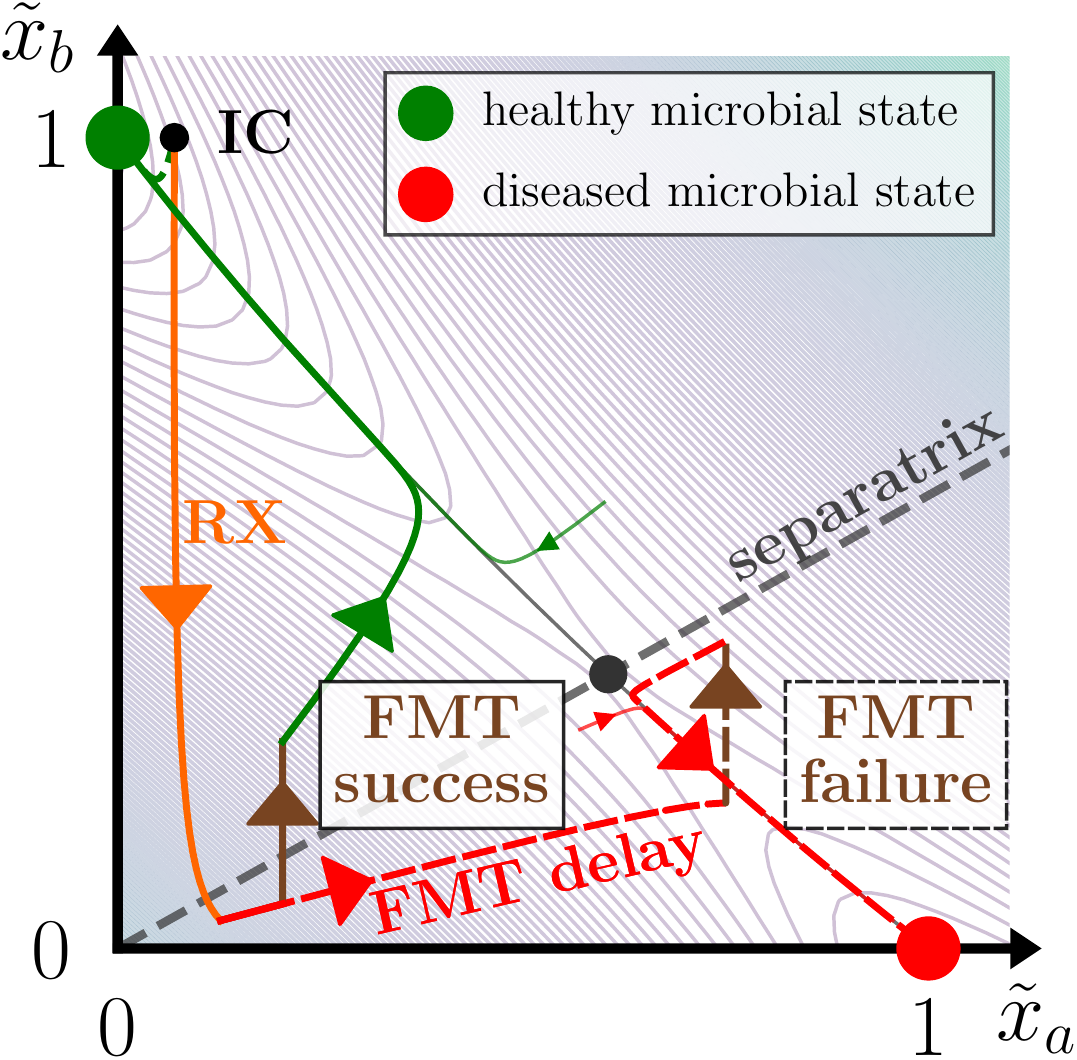}
\caption{{\bf The success of direct interventions depends on the size and timing of the transplant.}
We consider a clinically-inspired scenario that parallels antibiotic-induced CDI in the 2D gLV system Eq.~(\ref{eq:gLV_2D}).
First, a health-prone initial condition (IC) is depleted by antibiotics (RX, orange).
If a direct transplant (FMT, brown) is administered shortly after the antibiotics, the treatment steers the composition to a healthy state (FMT success).
Also, if the direct transplant is large enough to cross the separatrix, the intervention will be successful. 
Alternatively, if FMT administration is delayed and the transplant is too small, the microbial trajectory will instead attain the diseased state (FMT failure).
The basins of attraction of the healthy and diseased steady states are delineated by the separatrix, and the light contours depict isoclines of the potential energy landscape (given by a split Lyapunov function). 
This figure and caption are adapted from \cite{JonesCarlson2019}. 
\label{fig:transplant}}
\end{center} 
\end{figure}
%%%%%%%%%%%%%%%%%%%%%%%%%%%%%%%%%%%%%%%%%%%%%%%%%%%%%%%%

%%%%%%%%%%%%%%%%%%%%%%%%%%%%%%%%%%%%%%%%%%%%%%%%%%%%%%%%
\subsection{Transplant size and timing affect the efficacy of direct interventions}
%%%%%%%%%%%%%%%%%%%%%%%%%%%%%%%%%%%%%%%%%%%%%%%%%%%%%%%%
It is difficult to discern the influence of transplant size and timing on transplant efficacy in high-dimensional gLV systems, since only numerical methods are available to probe this relationship.
However, SSR provides a link between high-dimensional and reduced 2D systems, allowing the mechanisms that underlie direct interventions in high-dimensional systems to be understood in terms of their low-dimensional counterparts.

Fig.~\ref{fig:transplant} demonstrates how variability in transplant size and timing affects steady-state outcomes in a clinically-motivated scenario of CDI in a 2D gLV system.
Here, an initial condition is depleted by antibiotics (RX) to the extent that it enters a diseased basin of attraction.
Immediately, the reduced system makes clear that the location of the separatrix is crucial in determining the outcome of a microbial state: successful direct interventions must cross the separatrix.
The transplant size required for success depends on the composition of the transplant (in this figure transplants are composed entirely of $\tilde x_b$).
Furthermore, the required transplant size is variable, and depends on the ecological dynamics of the system--- in Fig.~\ref{fig:transplant} a transplant of the same size and composition is administered at two different times, but the later transplant is unsuccessful.

SSR allows for the relationships between transplant size, composition, and timing to be examined analytically in terms of an intuitive low-dimensional system.
Next, having characterized the operation of direct transplants in a two-dimensional system, we examine how direct transplants may be used to transition between steady states in the full context of a multistable gLV system.

%%%%%%%%%%%%%%%%%%%%%%%%%%%%%%%%%%%%%%%%%%%%%%%%%%%%%%%%
\section{Results}
%%%%%%%%%%%%%%%%%%%%%%%%%%%%%%%%%%%%%%%%%%%%%%%%%%%%%%%%

%%%%%%%%%%%%%%%%%%%%%%%%%%%%%%%%%%%%%%%%%%%%%%%%%%%%%%%%
\subsection{Transplant compositions and their success rates}
%%%%%%%%%%%%%%%%%%%%%%%%%%%%%%%%%%%%%%%%%%%%%%%%%%%%%%%%

\begin{figure}[t]
\begin{center}
\includegraphics[width=.65\textwidth]{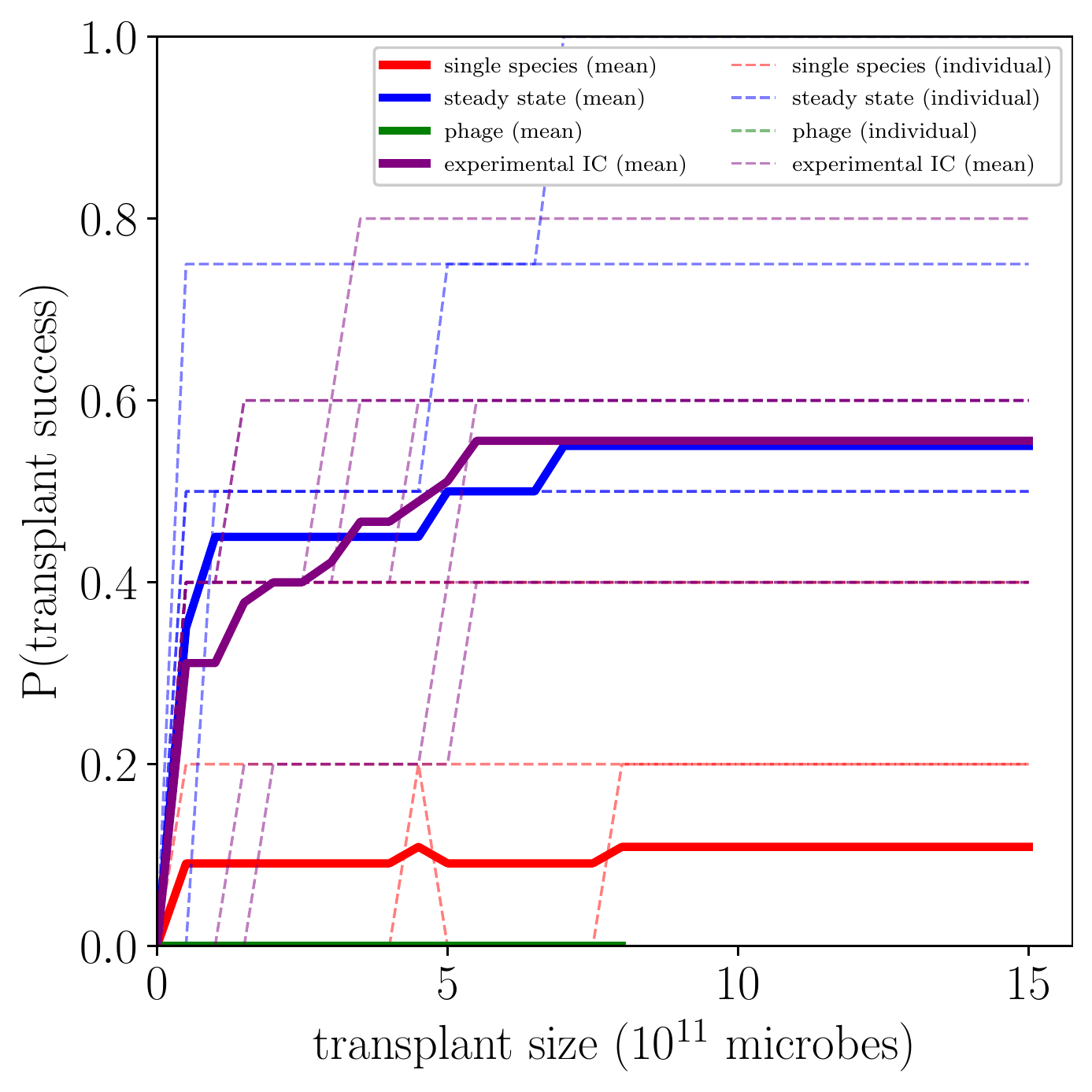}
\caption{{\bf Success rates of different microbial interventions at altering steady-state outcomes.}
Initial conditions were located one of the five steady states of the CDI model (displayed in Fig.~\ref{fig:reachable}) that were reached from experimentally-measured initial conditions.
Then these initial conditions were subjected to the introduction of a foreign population $\textbf{v}$ over a range of sizes $\lVert \textbf{v} \rVert_1$ (as shown on the x-axis).
If the introduced foreign population drove the initial condition into a different basin of attraction, the intervention was considered ``successful.''
For each of the four types of intervention, several candidate transplants $\textbf{v}$ were implemented, and the success rates of each candidate intervention were plotted (dashed lines).
A bulk success rate for each type of intervention (bold lines) was produced by averaging the success rates of the candidate transplants for each intervention type.
Details about the candidate transplant compositions used for each type of intervention are described in the main text.
\label{fig:transplant_success}}
\end{center} 
\end{figure}

\new{
Having demonstrated that transplants are capable of shifting an ecological state
into a different basin of attraction, we now investigate how
transplant composition influences transplant efficacy.
}
%The ecological state of a system can be permanently altered by deliberately introducing some foreign microbial composition or by selectively depleting the population of a single species.
%Fig.~\ref{fig:reachable_ss} provides proof-of-concept that an ecological trajectory can be shifted from one basin of attraction to another by introducing a foreign microbial population $\textbf{v}$, and Fig.~\ref{fig:transplant} demonstrates the same phenomenon graphically in a reduced 2D gLV system.
%In this section, the influence of transplant composition on direct intervention efficacy is examined using the high-dimensional CDI model.
%Different types of direct interventions vary in their ability to switch a trajectory between basins of attraction.
Three intervention types are considered: the introduction of a single-species ``probiotic,'' a stable community via ``fecal microbiota tranplantation'' (FMT), or the elimination of a single species via ``phage therapy.''
These interventions act as {\em in silico} proxies for medical therapies.
The success rates of these interventions are plotted as a function of the intervention magnitudes in Fig.~\ref{fig:transplant_success}.
As an example, the transplant administered in Fig.~\ref{fig:reachable_ss}d is considered a ``successful'' intervention, since it was able to alter a trajectory tending towards steady state D (the CD-infected steady state) and drive it towards steady state C (the healthy steady state).

%uses a transplant from the relatively stable microbiome of a healthy donor to cure dysbiotic microbiomes; FMT is 
Fecal microbiota transplantation is implemented in the model by setting the transplant $\textbf{v}$ proportional to a steady state of the gLV system.
In particular, transplants $\textbf{v}$ are set proportional to one of the five reachable steady states depicted in Fig.~\ref{fig:reachable}, or they are set proportional to one of the nine experimentally-measured initial conditions from the mouse experiment performed by Buffie {\em et al.} \cite{BuffiePamer2012}.
Probiotics are realized by setting the transplant $\textbf{v}$ proportional to a single microbial species.
Lastly, phage therapies are described by making the transplant $\textbf{v}$ negative, and setting it proportional to a single microbial species (so that it depletes a particular species).

In Fig.~\ref{fig:transplant_success} these interventions are applied to initial conditions located at the reachable steady states of the CDI model depicted in Fig.~\ref{fig:reachable}.
Interventions are considered successful when they shift the basin of
attraction an initial condition is in.
%For each type of intervention, several transplants are considered.
For the single species ``probiotic,'' transplants are solely composed of one of the 11 bacterial species of the model. 
The steady state ``FMT'' populations are composed of one of the five reachable steady states.
There are 11 ``phage'' interventions that each deplete a single bacterial species.
Lastly, the nine ``experimental IC'' foreign populations are proportional to the experimental initial conditions measured by Buffie {\em et al.} \cite{BuffiePamer2012}.
In Fig.~\ref{fig:transplant_success} the success rates of each intervention are plotted according to their magnitude (i.e. the one-norm $\lVert \textbf{v} \rVert_1$ of each foreign population).
For scale, the five reachable steady states of the CDI model vary in size between $3\times 10^{11}$ and $24 \times 10^{11}$ microbes, which informs the range of direct intervention sizes considered here.

As shown in Fig.~\ref{fig:transplant_success}, the success rates of multi-species interventions (steady states and experimental ICs) are significantly higher than the single-species interventions (single-species and phage therapies).
In particular, phage therapies are completely ineffective at altering the basin of attraction that an initial condition is in.
For each type of intervention, the success rates of each candidate transplant within each intervention type are plotted in dashed lines.
The bulk success rates of each intervention are plotted in bold, and are computed by averaging the success rates of the individual introduced foreign populations within each intervention type.

%%%%%%%%%%%%%%%%%%%%%%%%%%%%%%%%%%%%%%%%
%%%%%%%%%%%%%%%%%%%%%%%%%%%%%%%%%%%%%%%%

%%%%%%%%%%%%%%%%%%%%%%%%%%%%%%%%%%%%%%%%
\begin{figure}[t]
\begin{center}
\includegraphics[width=.8\textwidth]{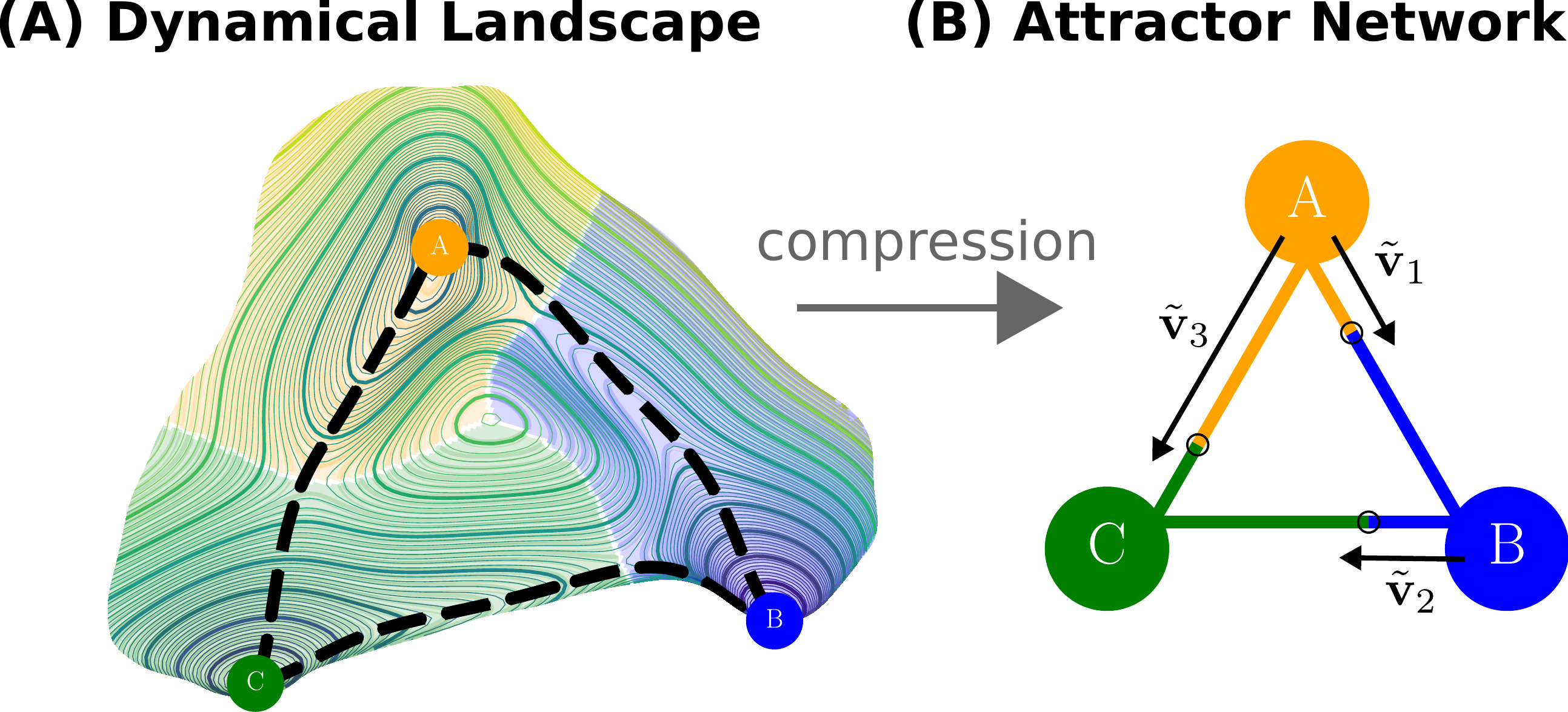}   
\caption{
\textbf{Schematic of a dynamical landscape and its corresponding attractor network.}
(A) Ecological dynamics that are not periodic or chaotic may be represented as particles flowing on a dynamical landscape.
Here, this artificial landscape exhibits three stable steady states at A (yellow), B (blue), and C (green).
The basins of attraction are colored according to the color of their associated steady state, and isoclines of the landscape are also plotted.
The black dotted line displays the value of the energy landscape along convex combinations of steady-state pairs.
(B) To compress this dynamical landscape into an attractor network, a graph is made in which nodes are steady states and edges are convex combinations of those steady states.
The edge color for a particular convex combination corresponds to the basin of attraction of that ecological state.
The small black circles in the attractor network correspond to the separatrices of the dynamical landscape.
The transplants $\tilde{\vec v}_1$, $\tilde{\vec v}_2$, and $\tilde{\vec v}_3$,
    whose compositions are given in the main text, demonstrate how direct
    interventions can alter the basin of attraction a system is in.
\label{fig:landscape}}
\end{center}
\end{figure}
%%%%%%%%%%%%%%%%%%%%%%%%%%%%%%%%%%%%%%%%

%%%%%%%%%%%%%%%%%%%%%%%%%%%%%%%%%%%%%%%%
\subsection{The dynamical landscape of a gLV system} 
%%%%%%%%%%%%%%%%%%%%%%%%%%%%%%%%%%%%%%%%

The ecological dynamics of gLV systems are dictated by complex feedbacks between populations.
%When these interactions are competitive $(M_{ij} < 0)$, there exist relaxationary dynamics that drive an initial microbial state towards a static equilibrium.
Conceptually, in the absence of periodic or chaotic dynamics these ecological dynamics can be interpreted as flowing downhill a dynamical landscape (e.g. a Lyapunov function) towards a point attractor.
%The dynamical landscape of a gLV system is determined by its growth rates and interaction parameters.
A visualization of a dynamical landscape for a two-dimensional state space is displayed schematically in Fig.~\ref{fig:landscape}a.
Here, the stable steady states A (yellow), B (blue), and C (green) are located at the minima of the landscape. 
Basins of attraction are displayed topographically and are the same color as their associated steady state.

In the CDI model, as demonstrated in Fig.~\ref{fig:reachable_ss} three distinct steady states can be reached from the same initial condition by administering different interventions (antibiotic administration and CD exposure).
In earlier work we showed that the transition between these steady states is sudden as a function of the magnitude of the intervention (for example, an antibiotic dose of 0.70 units leads to steady state C, while a dose of 0.72 units leads to steady state E) \cite{jones2018silico}.
Thus, the basins of attraction of these three steady states are touching, but the structure of this high-dimensional dynamical landscape is difficult to visualize.
To rectify this, attractor networks can be used to compress information about the basins of attraction of the high-dimensional phase space into a visually-digestible form.
Attractor networks were originally introduced by Wang {\em et al.} to study the controllability of gene regulatory networks associated with cancer \cite{WangLai2016}.

%Since gLV systems typically exhibit many such equilibria, the global landscape that ``stitches together'' the basins of attraction of many equilibria is considered, such as in Fig. 1.

In the schematic Fig.~\ref{fig:landscape}, panel (B) displays the attractor network for the dynamical landscape in panel (A).
In the attractor network, nodes are steady states of the high-dimensional system, edges are convex combinations of pairs of steady states, and the colors along edges correspond to the basins of attraction along each convex combination.

Attractor networks are especially valuable for mapping the landscape near a few steady states in a high-dimensional system.
For example, Fig.~\ref{fig:attractor_network} presents an attractor network for the five reachable steady states of the CDI model (described in Fig.~\ref{fig:reachable}).
The attractor network preserves geometric information about the basins of attraction of these five steady states, without requiring awkward visualization of an 11-dimensional state space.

The realization of antibiotic-induced CDI, as demonstrated in Fig.~\ref{fig:reachable_ss}, indicated that (i) antibiotic administration shifted a microbial trajectory from the healthy steady state C towards the antibiotic-depleted steady state E, (ii) antibiotic exposure coupled with CD inoculation caused the microbial trajectory to flow towards the CD-infected steady state D, and (iii) trajectories in the basin of attraction of the CD-infected steady state D could be driven to the healthy steady state C through the introduction of a foreign population.
The attractor network provides a compressed description of how an ecological state responds to external interventions, and complements and strengthens the numerical proof-of-concept analysis of antibiotic-induced CDI in Fig.~\ref{fig:reachable_ss}. 

%%%%%%%%%%%%%%%%%%%%%%%%%%%%%%%%%%%%%%%%
\subsection{Efficient navigation of an attractor network}
%%%%%%%%%%%%%%%%%%%%%%%%%%%%%%%%%%%%%%%%

%%%%%%%%%%%%%%%%%%%%%%%%%%%%%%%%%%%%%%%%
\begin{figure}[t]
\begin{center}
\includegraphics[width=.6\textwidth]{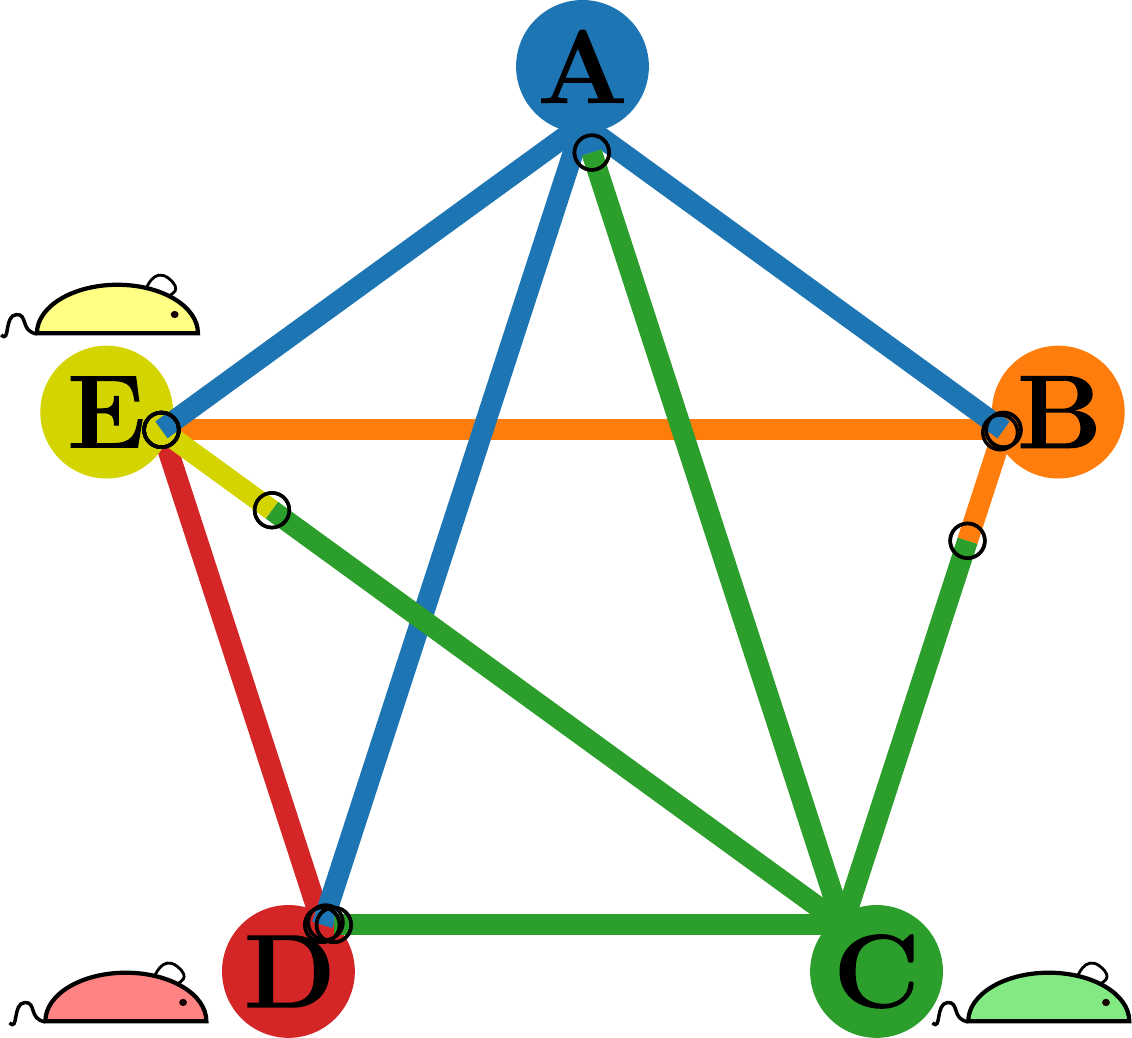}
\caption{\textbf{Attractor network of the five reachable steady states in the CDI model.}
The small black circles on each edge represent the numerically-calculated separatrices of each convex combination of steady states, which denote the boundary of the basins of attraction for a pair of steady states.
The relationships between steady states C (healthy), D (CD-infected), and E (antibiotic-depleted) are explored numerically in Fig.~\ref{fig:reachable_ss}.
Representing the basins of attraction of these reachable steady states as an attractor network allows for an intuitive low-dimensional description of the dynamical landscape of the high-dimensional CDI system.
\label{fig:attractor_network}}
\end{center} 
\end{figure}
%%%%%%%%%%%%%%%%%%%%%%%%%%%%%%%%%%%%%%%%

%%%%%%%%%%%%%%%%%%%%%%%%%%%%%%%%%%%%%%%%
\begin{figure}[t]
\begin{center}
\includegraphics[width=.9\textwidth]{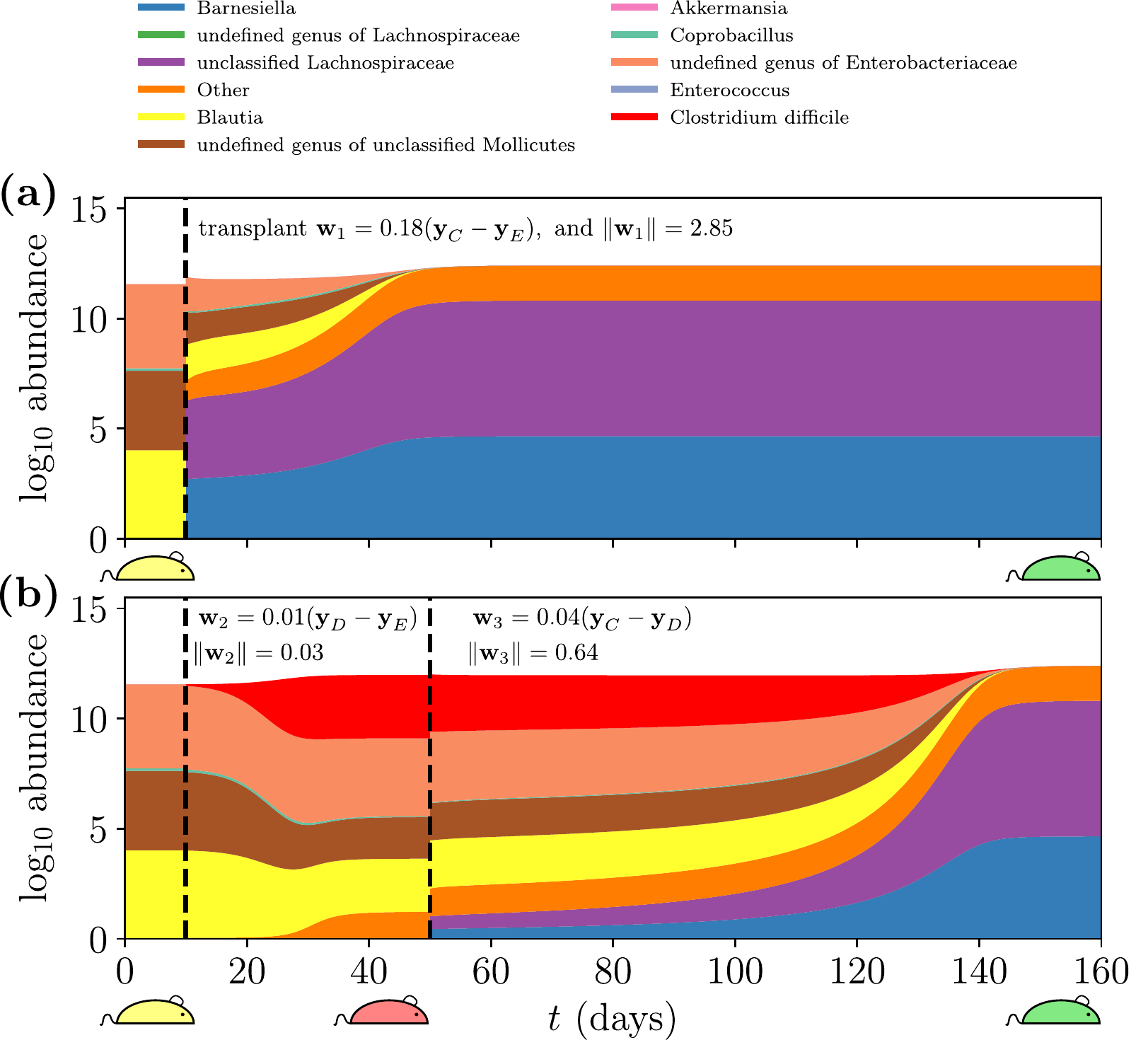}
\caption{\textbf{Direct interventions drive a microbial composition at steady state E (antibiotic-depleted, yellow mouse) towards steady state C (healthy, green mouse).}
%The microbial trajectories are displayed as in Fig.~\ref{fig:reachable_ss}.
(a) The microbial trajectory associated with a control protocol that drives the system directly from steady state E to steady state C.
This control protocol introduces a transplant $\vec w_1 = 0.18(\vec y_C - \vec y_E)$ on day 10.
The introduced transplant is of size $\lVert \vec w_1 \rVert = 2.85 \times 10^{11}$ microbes.
(b) A circuitous control protocol drives the system at steady state E first towards steady state D (CD-infected, red mouse), and then applies a second transplant to drive the system to steady state D.
By administering two smaller transplants sequentially, this control protocol requires fewer total microbes.
    requires a smaller fewer total microbes.
The protocol in (b) requires a cumulative transplant size that is 23\% the size of the protocol in (a).
\label{fig:transplant_time_course}}
\end{center} 
\end{figure}
%%%%%%%%%%%%%%%%%%%%%%%%%%%%%%%%%%%%%%%%

%%%%%%%%%%%%%%%%%%%%%%%%%%%%%%%%%%%%%%%%
\begin{figure}[t]
\begin{center}
\includegraphics[width=.6\textwidth]{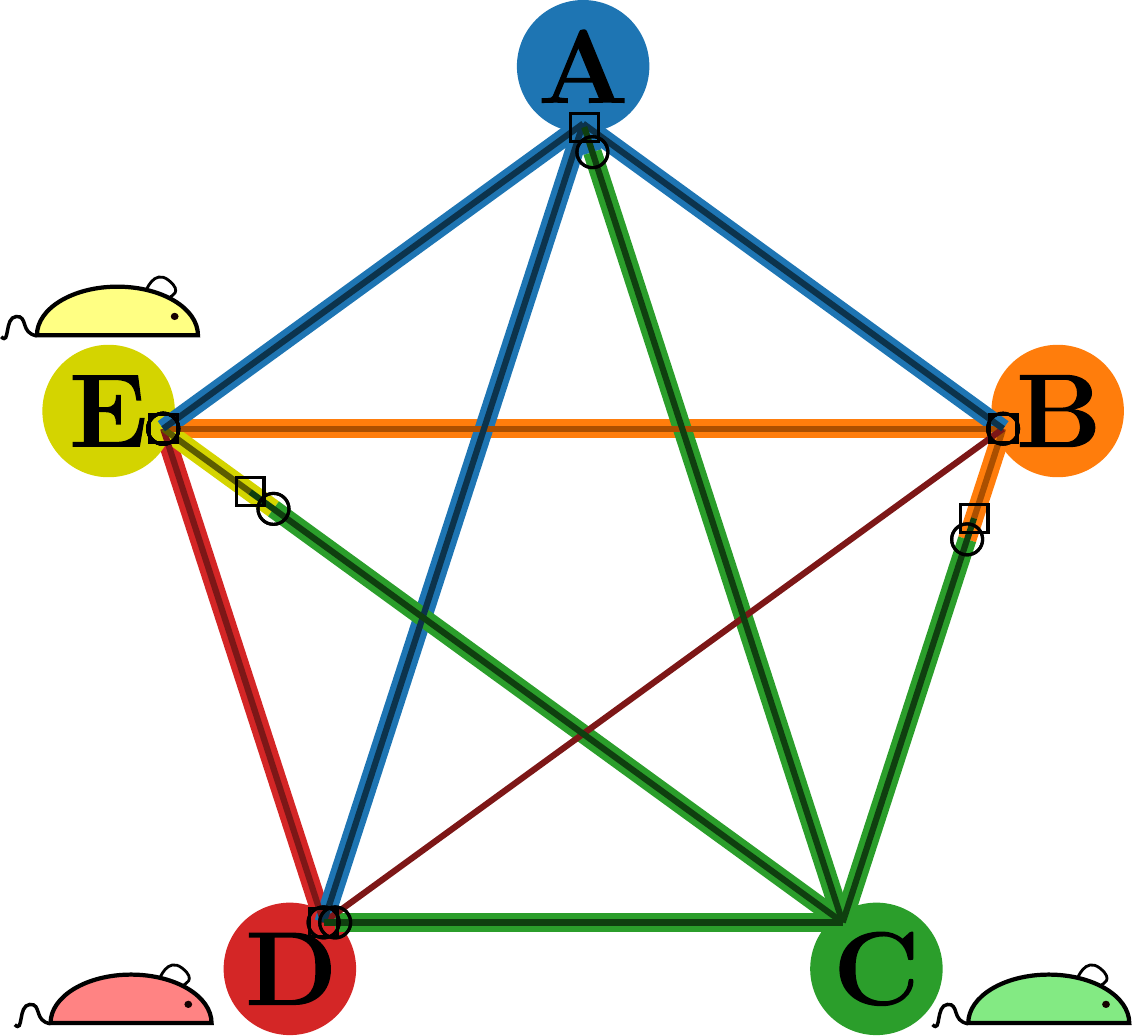}
\caption{\textbf{Attractor network generated exactly (numerically) and approximately (with SSR).}
This figure is identical to Fig.~\ref{fig:attractor_network}, but in addition
    to the small black circles corresponding to the numerically-calculated separatrix, the small black squares correspond to the separatrix locations as predicted by SSR.
SSR incorrectly predicts that any convex combination of steady states B and D will evolve towards steady state D, when in fact any convex combination of these steady states will evolve towards steady state A.
Despite this inaccuracy, SSR effectively and efficiently approximates the geometry of the dynamical landscape of the CDI gLV system. 
\label{fig:attractor_network_SSR}}
\end{center} 
\end{figure}
%%%%%%%%%%%%%%%%%%%%%%%%%%%%%%%%%%%%%%%%

The attractor network acts as a ``roadmap'' for a dynamical landscape, places microbial trajectories in the context of other nearby attractors, and can guide the administration of interventions in a gLV model.
In discussing how to navigate gLV systems, we will assume that we are able to drive the state of the system along convex combinations of pairs of steady states.

Explicitly, represent steady states A, B, and C in Fig.~\ref{fig:landscape} by the vectors $\vec x_A$, $\vec x_B$, and $\vec x_C$, respectively.
Then, starting from steady state $i$, a direct intervention in the direction of
steady states $j$ corresponds to a transplant $\tilde{ \vec v}=p(\vec x_j -
\vec x_i)$ in the gLV equations Eq.~(\ref{eq:transplant}), where $p$ varies
between 0 and 1 and describes the magnitude of the intervention.
For example, in the schematic attractor network Fig.~\ref{fig:landscape}b, starting from steady state A an introduced transplant $\tilde{\vec v}_1 = 0.3(\vec x_B - \vec x_A)$ will drive the state of the system into the basin of attraction of steady state B.

Explicitly, we seek to minimize the size of the intervention required to drive a system from an initial steady state into the basin of attraction of a target steady state $\lVert \tilde{\vec v} \rVert$.  
Using the same schematic attractor network from Fig.~\ref{fig:landscape}b, driving an initial condition at steady state B into the basin of attraction of steady state C will require a transplant $\tilde{\vec v}_2 = 0.3(\vec x_C - \vec x_B)$.
In this schematic the distances between each pair of steady states is defined
to be one unit, so applying these two interventions in a sequential manner will
drive an initial condition at steady state A into the basin of attraction of steady state C
with a total intervention size of 0.6.
Alternatively, to drive an initial condition at steady state A into the basin
of attraction of steady state C directly requires a transplant $\tilde{\vec v}_3 = 0.7(\vec x_C - \vec x_A)$, with a total intervention size of 0.7.
Thus, constructing an attractor network demonstrates that inherent ecological dynamics can be leveraged to efficiently control the state of an ecological system.

In the CDI model, the attractor network in Fig.~\ref{fig:attractor_network}
reveals that scenarios exist in which the most efficient control method follows indirect paths.
In Fig.~\ref{fig:transplant_time_course} the microbial time courses associated
with two control protocols--- one that follows a direct path, and another
that follows a circuitous path--- are compared.

Explicitly, let steady state $i$ of the CDI model correspond to the vector $\vec y_i$.
To transition from the antibiotic-depleted steady state E directly to the healthy steady state C requires a transplant $\vec v_1 = 0.162(\vec y_C - \vec y_E)$, with total transplant size $\lVert v_3 \rVert = 2.576$, where $\lVert \cdot \rVert$ is the 2-norm and each transplant is in units of $10^{11}$ microbes.
A similar control protocol is plotted in Fig.~\ref{fig:transplant_time_course}a, which successfully drives the state of the system into the basin of attraction of steady state C (the plotted transplant $\vec w_1$ is chosen to be slightly larger than the required transplant $\vec v_1$ for demonstrative purposes).

However, in this CDI model it is most efficient to apply two sequential interventions $\vec v_2$ and $\vec v_3$: the first transplant drives the system towards the CD-infected steady state D, $\vec v_2 = 0.001(\vec y_D - \vec y_E)$; and the second transplant drives the system towards the healthy steady state C, $\vec v_3 = 0.022(\vec y_C - \vec y_D)$.
Here, since steady state E is unstable in the direction of steady state D, an infinitesimal transplant is all that is needed (though for practical purposes we use a value of 0.001).
The sizes of these two sequential transplants are $\lVert v_2 \rVert = 0.0001$ and $\lVert v_3 \rVert = 0.360$.
A similar sequential control protocol is demonstrated in Fig.~\ref{fig:transplant_time_course}b; once again, for demonstrative purposes the transplants $\vec w_2$ and $\vec w_3$ are chosen to be slightly larger than the required transplants $\vec v_2$ and $\vec v_3$. 

Taken together, the circuitous control protocol E $\to$ D $\to$ C (total
transplant size $0.36$)  requires a smaller total intervention than the direct
path E $\to$ C (total transplant size $2.57$).
In general, taking advantage of this ``ecological inertia'' may reduce the magnitude of the intervention required to drive the system towards a target state.

%%%%%%%%%%%%%%%%%%%%%%%%%%%%%%%%%%%%%%%%
\subsection{Efficient construction of attractor networks with steady-state reduction}
%%%%%%%%%%%%%%%%%%%%%%%%%%%%%%%%%%%%%%%%

Finally, attractor networks can be approximated in constant time (in algorithmic complexity terms) with the dimensionality-reduction technique SSR.
The attractor network for the CDI model in Fig.~\ref{fig:attractor_network} was generated numerically by simulating a set of microbial trajectories, each originating at initial conditions along convex combinations of steady-state pairs, and then tracking the steady-state outcome of each simulation.
This procedure is computationally expensive even when bisection-type algorithms are implemented to identify the location of the separatrix. 

As derived in previous work, SSR allows for the separatrix of the approximate SSR-reduced 2D gLV system to be computed analytically in a Taylor series about a semi-stable fixed point $(x_a^*, x_b^*)$ of the reduced system \cite{JonesCarlson2019}.
For example, the separatrix of the 2D gLV system in Fig.~\ref{fig:transplant} was created in this fashion.
Though the current attractor network is relatively small, consisting of five
nodes and ${5 \choose 2}$ edges, scaling up the size of a numerically-calculated attractor network quickly becomes computationally infeasible.
On the other and, larger attractor networks can be efficiently approximated with SSR: as a demonstration of the accuracy with which SSR captures the location of the separatrices, in Fig.~\ref{fig:attractor_network_SSR} the squares on each edge represent the separatrix predicted by applying SSR to the corresponding steady-state pair.

The SSR-generated attractor network incorrectly predicts that convex combinations of steady states B and D will flow towards steady state D, when in fact they will flow towards steady state A.
This occurs because the pairwise connections in the graph representation of the
attractor network may oversimplify the complex topography of the original
landscape.
Still,  the strong agreement between the exact and SSR-generated separatrices indicates the potential of SSR to efficiently generate large attractor networks to first-order that intuitively describe relevant ecological dynamics.

\section{Discussion}

\new{
In this section we provide additional biological, ecological, and mathematical
context for the microbiome control problem considered in this paper. First,
Section \ref{subsec:microbiome} discusses biological evidence relating
microbiome composition to host health, as well as clinical evidence for the
success of FMT in treating CDI. Then, Section \ref{subsec:modeling} surveys
different classes of ecological models (of varying levels of complexity) that
have been used to model the microbiome. Finally, Section \ref{subsec:control}
considers techniques from nonlinear control theory that might be applicable
to gLV systems in future work. 
}

\subsection{Microbiome composition is associated with host health}\label{subsec:microbiome}

Microbiome research has been the recent beneficiary of advances in experimental 16S pyrosequencing techniques that have revealed similarities in the microbiome compositions of people suffering various diseases, which differ from the microbiome compositions of healthy individuals \cite{shreiner2015gut}.
These disease-associated microbiome compositions are called {\em dysbiotic}, and are observed in people that suffer cardiovascular disease, ulcerative colitis, and irritable bowel disease \cite{lee2010has, shreiner2015gut}.
%Damage to the microbiome has severe implications for the host.
%For example, in the subgingival microbiome and the gut microbiome several pathogenic species of bacteria were found only in hosts who suffered from periodontitis and CDI \cite{paster2001bacterial,schubert2014microbiome}.

In this chapter we studied a mathematical model of {\em C. difficile} infection (CDI). 
CDI occurs when the bacteria \textit{C. difficile} (CD) colonizes the gut and becomes sufficiently abundant to induce colon inflammation and diarrhea via the production of toxins TcdA and TcdB \cite{voth2005clostridium}.
This disease is an explicit example of how impaired microbiome compositions cause disease.

Traditional treatments for CDI seek to eliminate the presence of CD by administering antibiotics like vancomycin or metronidazole \cite{mcfarland2002breaking}.
However, CDI recurrence rates that range from 30-65\% and concerns about antibiotic resistance have heralded fecal microbiota transplantation (FMT) as an alternative treatment for CDI \cite{brandt2012long,spigaglia2011multidrug}.
FMT is a bacteriotherapy that attempts to alter the composition of a dysbiotic microbiome (e.g., the microbiome of a person suffering CDI) by engrafting a foreign microbial population provided by a healthy donor into it.
When successful, FMT causes the restoration of a healthy microbiome, which suppresses the growth of CD as well as the production of CDI-associated toxins \cite{khoruts2016understanding}.
The success rate of FMT for treating CDI approaches 90\% \cite{brandt2012long}.

FMT has been proposed as a treatment for other gastointestinal diseases including irritable bowel syndrome (IBS), inflammatory bowel disease, ulcerative colitis, and  Crohn's disease. 
Recent clinical trials have returned promising but inconclusive evidence regarding the success of FMT in treating these conditions \cite{coryell2018gut,franzosa2018gut,de2013transient,anderson2012systematic}.
For other conditions with distinct microbial signatures--- for example, autism spectrum disorder (ASD) or metabolic syndrome \cite{vuong2017emerging,perry2016acetate} --- a more intrinsic question regarding the efficacy of bacteriotherapies remains: can symptoms of these conditions be ameliorated by altering the gut microbiome composition? 
For ASD, preliminary evidence supports this hypothesis.
Patients with ASD were given Microbiota Transfer Therapy, a treatment that consists of an initial course of antibiotics followed by regular FMT treatments for ten weeks, and this treatmenet caused lasting shifts to their microbiome compositions and reductions in their ASD symptoms \cite{kang2019long}. 

At this time, the ability of FMT to treat microbiome-associated diseases is variable.
FMT appears to be an effective treatment for CDI, but recent clinical studies have been unable to conclusively show that FMT is an effective treatment for IBS \cite{xu2019efficacy}.
Part of this discrepancy might be due to antibiotics, which are typically administered to CDI patients before FMT is attempted.
More generally, the factors that contribute to the success or failure of FMT are not yet fully known.
To shed light on these factors, this chapter used mathematical models to examine the ecological dynamics that underlie FMT.

%%%%%%%%%%%%%%%%%%%%%%%%%%%%%%%%%%%%%%%%
\subsection{Advances in modeling the microbiome}\label{subsec:modeling}
%%%%%%%%%%%%%%%%%%%%%%%%%%%%%%%%%%%%%%%%

Models of the microbiome can span levels of organization, stochasticity, neutrality, and complexity.
Recent advances in multi-omics sequencing have buoyed microbiome modeling by measuring the abundances of the microbial genomes, mRNA transcripts, proteins, and metabolites contained in the microbiome \cite{biggs2015metabolic}.

For microbial ecosystems with well-characterized metabolic pathways, flux balance analysis (FBA) predicts the production and consumption of metabolites by microbes at steady state using inferred stochiometric matrices  \cite{biggs2015metabolic}.
In dynamic flux balance analysis (dFBA), FBA is generalized to track metabolite abundances over time, at the expense of requiring additional kinetic parameters in the model  \cite{biggs2015metabolic}.
In both of these metabolite-based analyses, fine-grained details about metabolite kinetics need to be either measured or fit. 
This required level of detail makes FBA and dFBA ill-suited for analyzing complex microbial ecosystems like the human microbiome which, for example, consists of $\sim$1000 microbial species and $\sim$100000 metabolites \cite{lloyd2016healthy,wishart2018drugbank}.

To account for the complexity present in real-world microbial systems, coarse-graining approaches have been employed to model aggregate microbial dynamcs and processes.
For example, in wastewater treatment activated sludge models (ASMs) track the abundance of only a few nutrients (e.g. organic carbon, nitrogen, and phosphorus) and the population dynamics of only a few aggregate microbial populations (e.g. microbes that consume carbon, nitrogen, and phosphorus) \cite{bucci2014towards}.
Though less detailed than FBA and dFBA models, ASMs require significantly fewer parameters and are able to provide relevant information regarding the design and function of wastewater treatment processes.

Generalized Lotka-Volterra (gLV) models, which are studied in this chapter, assume that the production and consumption of nutrients can be implicitly captured by pairwise interactions between microbial populations. 
Thus, gLV models only track microbial population abundances. 
Some gLV models (including the CDI model) assume that microbial species within a genus are indistinguishable, and study microbial dynamics coarse-grained at the genus level.
Part of the convenience of gLV models is that they may be parameterized with time-series microbial abundance data, which is readily available due to advances in genomic sequencing.
Several algorithms have been developed to infer growth rate and interaction parameters of gLV models from time-series abundance data (e.g. LIMITS \cite{fisher2014identifying, bucci2016mdsine}).
gLV models have been used to investigate microbial interactions in cheese \cite{mounier2008microbial}, to probe how antibiotic perturbations alter the gut microbiome in the context of CDI \cite{stein2013ecological}, and to inform treatment of polymicrobial urinary tract infections \cite{de2017interaction}.
When gLV models accurately approximate the underlying microbial systems, simulated interventions (e.g. antibiotic administration or fecal microbiota transplantation) can guide clinical efforts to alter the composition of diseased microbiomes \cite{stein2013ecological, buffie2015precision,jones2018silico}.

Each of these previously discussed models make the simplifying assumption that microbial processes and the resulting microbial dynamics are deterministic.
In ecology these deterministic approaches are typically used to model niche processes.
While microbial dynamics can be relatively consistent in some systems (e.g. in soil microbial communities \cite{friedman2017community}), variability--- reflecting neutral processes--- is generally evident in microbial communities (e.g. in human \cite{li2016testing} or fly \cite{obadia2017probabilistic} microbiomes).
Mathematical models have been developed to account for this variability \cite{faust2018signatures}.
For example, Sloan {\em et al.} proposed a model to predict microbial abundances based solely on birth-death processes and immigration following Hubbell's neutral theory \cite{sloan2006quantifying}, while the Ricker model has been used to introduce stochasticity into gLV models \cite{fisher2014identifying}.

\new{
    \subsection{Alternate approaches from nonlinear control
    theory}\label{subsec:control}
Existing 
theory describes the controllability of polynomial differential equations using 
differential geometry \cite{Kunita1979}, and these techniques should be applicable
to gLV systems. For example, the Pontryagin maximum principle
provides a necessary condition for whether a proposed control protocol is
optimal \cite{BoltyanskiyPontryagin1962}. This approach has been previously applied to a
optimal sampled-data control problem in the context of a biomechanical model that predicts how muscles respond to electrical
stimulations 
\cite{BourdinTrelat2016}. In this biomechanical study, a version of the
Pontryagin maximum principle specific to optimal sampled-data control problems
was used to identify the optimal timing of electrical
stimulations that maximize the muscular force response \cite{BakirRouot2020}.
Since our microbiome control problem also treats direct interventions as Dirac
delta inputs, this modified version of the Pontryagin
maximum principle could be applied to gLV systems. 
}

\section{Conclusion}

The ability to control steady-state outcomes of ecological systems has a broad practical appeal.
These control protocols will rely on a foundation of ecological theory that is still under exploration.
Here we introduce a technique to map the dynamical landscape of gLV systems with attractor networks.
Although our analysis was solely concerned with gLV systems, fundamental ecological behaviors are demonstrated. 
For example, when attempting to control the steady-state outcome of a gLV system, the relevant objective is driving the system into the target basin of attraction rather than exactly driving the system to the target steady state. 
Additionally, when attempting to drive an ecosystem towards a target state, it might be more efficient to use a multi-step control protocol.
With the steady progression of ecological theory it is feasible that precision
bacteriotherapies, based upon ecological models of the microbiome, will one day
become a commonplace medicine for the microbiome.

\subsubsection*{Acknowledgements}
This material was based upon work supported by the National Science Foundation Graduate Research Fellowship Program under Grant No.~1650114. Any opinions, findings, and conclusions or recommendations expressed in this material are those of the author(s) and do not necessarily reflect the views of the National Science Foundation. This work was also supported by the David and Lucile Packard Foundation and the Institute for Collaborative Biotechnologies through Contract No. W911NF-09-D-0001 from the U.S. Army Research Office.  The funders had no role in study design, data collection and analysis, decision to publish, or preparation of the manuscript.

\appendix

\clearpage

\section*{\new{Appendix A: Parameters of the 11D CDI gLV model}}

\newcommand{\colwidth}{28mm}
\def\rot{\rotatebox}
{
\begin{table}[h]
\begin{center}
\scriptsize
    \new{
    \begin{tabular}{ r|r| }
        \cline{2-2}
        Barnesiella & 0.3680\\ \cline{2-2}
        undefined genus of Lachnospiraceae&0.3102 \\ \cline{2-2}
        unclassified Lachnospiraceae&0.3561 \\ \cline{2-2}
        Other&0.5400 \\ \cline{2-2}
        Blautia&0.7089 \\ \cline{2-2}
        undefined genus of unclassified Mollicutes&0.4706 \\ \cline{2-2}
        Akkermansia&0.2297 \\ \cline{2-2}
        Coprobacillus&0.8300 \\ \cline{2-2}
        undefined genus of Enterobacteriaceae&0.3236 \\ \cline{2-2}
        Enterococcus&0.2907 \\ \cline{2-2}
        Clostridium difficile&0.3918 \\ \cline{2-2}
    \end{tabular}
}
\end{center}
    \caption{\new{Growth rates $\rho_i$ of each microbial population $i$ of the CDI
    model constructed by Stein {\em et al.} \cite{stein2013ecological}. 
    This table and caption are adapted from
    \cite{jones2018silico}.} \label{tab:mu} }
\end{table}
}

{
\begin{table}[h]
\begin{center}
\scriptsize
    \new{
    \begin{tabular}{ r|r|r|r|r|r|r|r|r|r|r|r| }
        \multicolumn{1}{c}{} &
        \multicolumn{1}{c}{\rot{90}{Barnesiella}}&
        \multicolumn{1}{c}{\rot{90}{und. Lachnospiraceae}}&
        \multicolumn{1}{c}{\rot{90}{uncl.  Lachnospiraceae}}&
        \multicolumn{1}{c}{\rot{90}{Other}}&
        \multicolumn{1}{c}{\rot{90}{Blautia}}&
        \multicolumn{1}{c}{\rot{90}{und. Mollicutes}}&
        \multicolumn{1}{c}{\rot{90}{Akkermansia}}&
        \multicolumn{1}{c}{\rot{90}{Coprobacillus}}&
        \multicolumn{1}{c}{\rot{90}{und.  Enterobacteriaceae}}&
        \multicolumn{1}{c}{\rot{90}{ Enterococcus}}&
        \multicolumn{1}{c}{\rot{90}{Clostridium difficile}}\\ \cline{2-12}
        Barnesiella & -0.205 & 0.098 & 0.167 & -0.164 & -0.143 & 0.019 & -0.515 & -0.391 & -0.268 & 0.008 & 0.346 \\ \cline{2-12}
        und. Lachno. &0.062 & -0.104 & -0.043 & -0.154 & -0.187 & 0.027 & -0.459 & -0.413 & -0.196 & 0.022 & 0.301 \\ \cline{2-12}
        uncl. Lachno. &0.143 & -0.192 & -0.101 & -0.139 & -0.165 & 0.013 & -0.504 & -0.772 & -0.206 & -0.006 & 0.292 \\ \cline{2-12}
        Other&0.224 & 0.138 & 0.000 & -0.831 & -0.223 & 0.220 & -0.205 & -1.009 & -0.400 & -0.039 & 0.666 \\ \cline{2-12}
        Blautia&-0.180 & -0.051 & -0.000 & -0.054 & -0.708 & 0.016 & -0.507 & 0.553 & 0.106 & 0.224 & 0.157 \\ \cline{2-12}
        und. Mollicutes &-0.111 & -0.037 & -0.042 & 0.041 & 0.261 & -0.422 & -0.185 & -0.432 & -0.264 & -0.061 & 0.164 \\ \cline{2-12}
        Akkermansia &-0.126 & -0.185 & -0.122 & 0.380 & 0.400 & -0.160 & -1.212 & 1.389 & -0.096 & 0.191 & -0.379 \\ \cline{2-12}
        Coprobacillus&-0.071 & 0.000 & 0.080 & -0.454 & -0.503 & 0.169 & -0.562 & -4.350 & -0.207 & -0.223 & 0.443 \\ \cline{2-12}
        und. Enterobac. &-0.374 & 0.278 & 0.248 & -0.168 & 0.084 & 0.033 & -0.232 & -0.395 & -0.384 & -0.038 & 0.314 \\ \cline{2-12}
        Enterococcus &-0.042 & -0.013 & 0.024 & -0.117 & -0.328 & 0.020 & 0.054 & -2.096 & 0.023 & -0.192 & 0.111 \\ \cline{2-12}
        C. difficile &-0.037 & -0.033 & -0.049 & -0.090 & -0.102 & 0.032 & -0.181 & -0.303 & -0.007 & 0.014 & -0.055 \\ \cline{2-12}
    \end{tabular}
    }
\end{center}
    \caption{\new{Interactions $K_{ij}$ between microbial populations $i$ and $j$
    of the CDI model constructed by Stein {\em et al.} \cite{stein2013ecological}. Each
    interaction $K_{ij}$ describes the effect of population $j$ on population
    $i$.  Population $i$ is given on the left, and population $j$ is given on
    the top. This table and caption are adapted from \cite{jones2018silico}.
    \label{tab:M}}}
\end{table}
}

\clearpage

\bibliographystyle{unsrt}
%\bibliography{Bib.bib}

\end{document}